\documentclass[preprint,showpacs,preprintnumbers,amsmath,amssymb,superscriptaddress]{revtex4-2}

\usepackage{bm,latexsym}
\usepackage{graphicx}
\usepackage{amsmath}
\usepackage{mathcomp}
\usepackage{amssymb}
\usepackage{mathrsfs} 
\usepackage{CJK}
\usepackage{ulem}
\usepackage{xcolor}
\usepackage{soul}
\usepackage{gensymb}

\begin{document}

\title{
Observation of the sliding  phason mode of the incommensurate magnetic texture in Fe/Ir(111)}

\author{Hung-Hsiang Yang}
\affiliation{Physikalisches Institut, Karlsruhe Institute of Technology, Wolfgang-Gaede Straße 1, 76131 Karlsruhe, Germany}

\author{Louise Desplat}
\affiliation{Universit\'e de Strasbourg, CNRS, Institut de Physique et Chimie des Mat\'eriaux de Strasbourg, UMR 7054, F-67000 Strasbourg, France}
\affiliation{Nanomat/Q-mat/CESAM, Universit\'e de Li\`ege, B-4000 Sart Tilman, Belgium}

\author{Volodymyr P. Kravchuk}
\affiliation{Institut f\"ur Theoretische Festkörperphysik, Karlsruhe Institute of Technology, D-76131 Karlsruhe, Germany}
\affiliation{Leibniz-Institut für Festkörper- und Werkstoffforschung, IFW Dresden, D-01171 Dresden, Germany}
\affiliation{Bogolyubov Institute for Theoretical Physics of the National Academy of Sciences of Ukraine, 03143 Kyiv, Ukraine}

\author{Marie Herv\'{e}}
\affiliation{Institut des NanoSciences de Paris, Sorbonne University and CNRS-UMR7588, 4 place Jussieu, 75005 Paris, France}
\author{Timofey Balashov}
\affiliation{II. Physikalisches Institut B, RWTH Aachen University, Otto-Blumenthal Straße, 52074 Aachen, Germany}

\author{Simon Gerber}
\affiliation{Physikalisches Institut, Karlsruhe Institute of Technology, Wolfgang-Gaede Straße 1, 76131 Karlsruhe, Germany}



\author{Markus Garst}
\affiliation{Institut f\"ur Theoretische Festkörperphysik, Karlsruhe Institute of Technology, D-76131 Karlsruhe, Germany}
\affiliation{Institute for Quantum Materials and Technology, Karlsruhe Institute of Technology, D-76021 Karlsruhe, Germany}

\author{Bertrand Dup\'e}
\affiliation{Fonds de la Recherche Scientifique (FNRS), B-1000 Bruxelles, Belgium}
\affiliation{Nanomat/Q-mat/CESAM, Universit\'e de Li\`ege, B-4000 Sart Tilman, Belgium}

\author{Wulf Wulfhekel}
\affiliation{Physikalisches Institut, Karlsruhe Institute of Technology, Wolfgang-Gaede Straße 1, 76131 Karlsruhe, Germany}
\affiliation{Institute for Quantum Materials and Technology, Karlsruhe Institute of Technology, D-76021 Karlsruhe, Germany}

\date{\today}

\begin{abstract}
The nanoscopic magnetic texture forming in a monolayer of iron on the (111) surface of iridium, Fe/Ir(111), is spatially modulated and uniaxially incommensurate with respect to the crystallographic periodicities. As a consequence, a low-energy magnetic excitation is expected that corresponds to the sliding of the texture along the incommensurate direction, i.e., a phason mode, which we explicitly confirm with atomistic spin simulations. Using scanning tunneling microscopy (STM), we succeed to observe this phason mode experimentally. It can be excited by the STM tip, which leads to a random telegraph noise in the tunneling current that we attribute to the presence of two minima in the phason potential due to the presence of disorder in our sample. This provides the prospect of a floating phase in cleaner samples and, potentially, a commensurate-incommensurate transition as a function of external control parameters.
\end{abstract}


\maketitle


\section{Introduction}

Incommensurate structures arise when two distinct periodic subsystems coexist but interact only weakly. In this case, their periodicities are, generically, not simply related; the ratio of their wavelength is an irrational number \cite{ChaikinLubensky}. Typical examples are periodic layers of atoms adsorbed on a crystalline substrate \cite{LNP}, spin- and charge density waves in solids \cite{Gruener} or intergrowth compounds \cite{VanSmaalen1995}. 
As a characteristic feature, incommensurate structures possess a gapless phason mode representing the sliding of the two periodic subsystems with respect to each other \cite{Overhauser1971}. This phason mode profoundly affects the system's properties:   
it stabilizes incommensurate structures giving rise to floating phases \cite{Bak1980}, it governs, in particular, the optical response of charge-density waves \cite{Gruener1988}, it causes strong-coupling superconductivity in bismuth at high pressures \cite{Brown2018}, and its properties are intimately connected, as shown recently, to the twist-angle variance in twisted bilayer graphene \cite{Ochoa2022}. However, if the interaction between the two subsystems increases, a commensurate-incommensurate (C-IC) transition takes place, where the periodicities become commensurate with the ratio of the respective wavelengths being a rational number. As a consequence, the phason mode acquires a gap \cite{Bruce1978}. 

Spatially modulated magnetic textures are a prime example for incommensurate phases as discussed early by Dzyaloshinskii \cite{Dzyaloshinsky1965}. At the focus of this work is the two-dimensional magnetic texture materializing in Fe/Ir(111), i.e., in a monolayer of iron on the (111) surface of iridium, that was discovered and characterized in the seminal works of von Bergmann  {\it et al.} \cite{vonBergmann2006} and Heinze {\it et al.} \cite{Heinze2011}. Iron forms a two-dimensional hexagonal lattice on Ir(111) with lattice constant $a = 2.715$ \AA\, whereas the magnetic texture is an approximate square lattice with lattice constant $A \approx 1$ nm. It was shown in Ref.~\cite{Heinze2011} that the strong Dzyaloshinskii-Moriya interaction at the Fe/Ir(111) interface and the four-spin interaction play a central role in stabilizing the magnetic modulation with a relatively small periodicity of only $A/a \approx 3.68$. 
The magnetic texture is thus nanoscopic \cite{Heinze2011} and far from the continuum limit as the rotation of spins from lattice site to lattice site is on the order of $360^\circ a/A \approx 100^\circ$. 
 
A system composed of two-dimensional hexagonal and square lattices, representing the arrangement of iron and the magnetic texture of Fe/Ir(111), respectively, is especially interesting because the two distinct symmetries preclude a  complete commensurability. A previous example known to us is the adsorption of Argon on the (100) surface of MgO \cite{LNP,Meichel1986}. Whereas the MgO substrate has square lattice symmetry the lateral interaction between adsorbed Ar favours an incommensurate hexagonal arrangement at high Ar densities. If the Ar density is reduced, two subsequent uniaxial C-IC transitions take place, where a commensurability is first attained along one $\langle 100\rangle$ axis and afterwards along the other axis, thereby changing the lattice symmetry from hexagonal to orthorhombic. 
 
Similarly, it seems that the symmetry of the magnetic texture in Fe/Ir(111) also deviates from that of a square lattice to enhance commensurability. Its periodicity  is, except at zero magnetic field, well resolved by spin-polarized scanning tunneling microscopy (STM) \cite{vonBergmann2006} indicating that the texture is not floating on the timescale of the STM experiment. It was pointed out in Ref.~\cite{vonBergmann2006} that a fully commensurate $3\times 5$ magnetic texture closely resembles the observations by STM implying a distortion from a square to a centered rectangular lattice with $A/a = \sqrt{13} \approx 3.61$, see Fig.~\ref{fig1}a. However, a careful analysis of second-order magnetic Bragg peaks obtained via scans of the tunnelling anisotropic magnetoresistance (TAMR) suggested a small deviation from this commensurability \cite{Heinze2011}. This poses the question whether the magnetic texture is fully or only partially incommensurate or whether different phases with distinct commensurabilities exist which are separated by uniaxial C-IC transitions, that can be controlled by temperature, magnetic field, or another parameter. 

In order to shed light on these questions, we investigate in the present work the properties of the sliding phason mode expected for an incommensurate magnetic texture in Fe/Ir(111). Carefully analyzing STM images of the magnetic texture, we identify a uniaxial incommensurability and the crystallographic orientation along which the phason mode is expected to be sliding the texture. This is  explicitly verified by atomistic spin simulations on a large unit cell that corresponds to a commensurate approximation of the magnetic order. These theoretical expectations are confirmed experimentally with our STM setup. The phason mode can be excited with the STM tip leading to random telegraph noise in the tunneling current, that we ascribe to the presence of two metastable minima in the phason potential due to the presence of disorder in our sample.

\section{Results}
\label{sec:RESULTS}

\subsection{Uniaxial incommensurability of the magnetic texture in Fe/Ir(111)}

\begin{figure}
\includegraphics[width=\columnwidth]{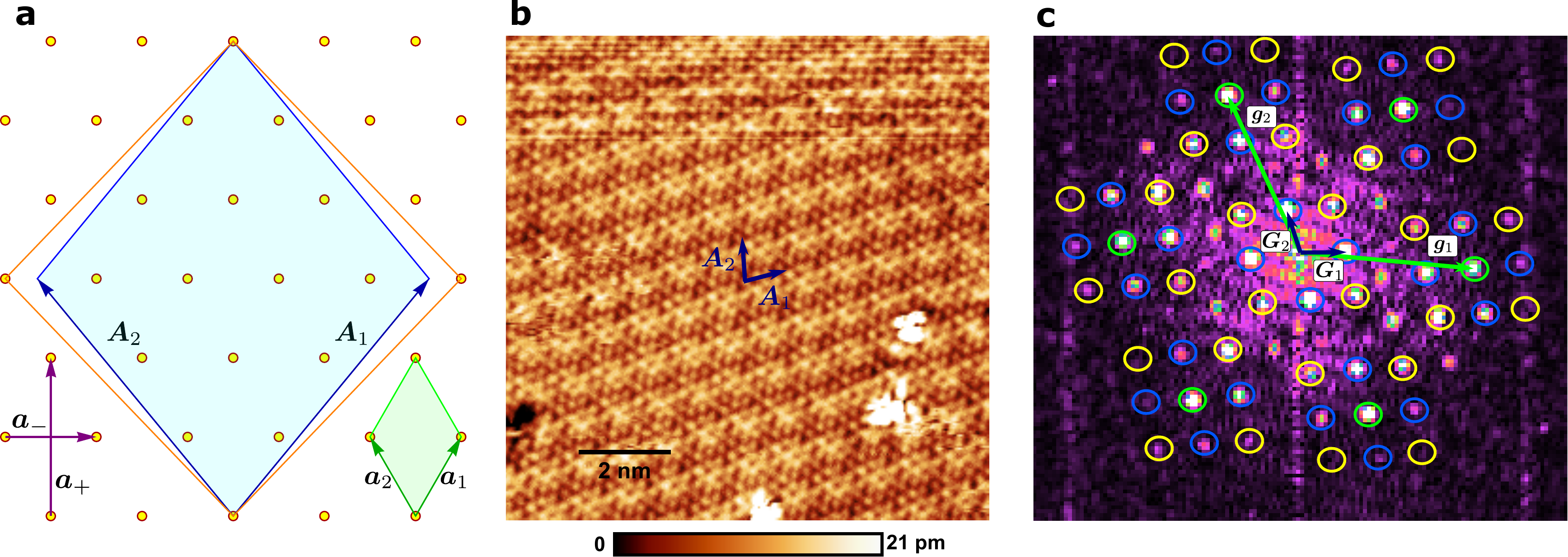}
\caption{{\bf a.} Atomic and magnetic Bravais lattices of Fe/Ir(111). Yellow dots represent the hexagonal lattice of Fe atoms, with a unit cell given by the green-shaded rhombus and primitive vectors ${\bf a}_1$ and ${\bf a}_2$, with ${\bf a}_\pm = {\bf a}_1 \pm {\bf a}_2$. 
The blue-shaded rhombus is a primitive unit cell of the magnetic order, with primitive vectors ${\bf A}_1$ and ${\bf A}_2$. The magnetic order possesses a uniaxial commensurability, ${\bf A}_1 + {\bf A}_2 = 3 {\bf a}_+ $. The orange lines represent the 
$3\times 5$ commensurate approximation of the magnetic unit cell.
%
{\bf b.} Atomically resolved spin-polarized STM image of the magnetic order ($U$=10 mV, $I$=15 nA) on Fe/Ir(111). The primitive vectors of the magnetic order of panel {\bf a} are also indicated. The topographic colour code is given in the legend.
{\bf c}. Fourier transform of the image in {\bf b}. Green circles and green arrows denote reciprocal lattice spots of Fe atoms and primitive vectors, respectively. Blue circles and blue arrows denote magnetic reciprocal lattice spots and primitive vectors, respectively, obtained via the TMR effect. Yellow circles are obtained via TAMR and correspond to higher-order magnetic Bragg peaks. The magnetic order possesses the commensurability  ${\bf g}_1 + {\bf g}_2 = 3 ({\bf G}_1 + {\bf G}_2)$. The corresponding lattices in real space are shown in panel {\bf a}.
}
\label{fig1}
\end{figure}

Figure \ref{fig1}b presents an atomically resolved STM topography of a single monolayer 
of Fe on Ir(111) recorded with a spin-polarized tip. The well-known magnetic texture of non-collinear spins \cite{Heinze2011} can clearly be seen in the topography as an approximate square lattice. The corresponding Fourier-transformed image shown in Fig.~\ref{fig1}c gives information on the magnetic unit cell with respect to the atomic crystal structure. The green circles represent atomic lattice spots that are well approximated by a hexagonal lattice. Its primitive reciprocal lattice vectors are shown by the green arrows ${\bf g}_1$ and ${\bf g}_2$. 
The blue and yellow circles mark magnetic spots due to the tunneling magnetoresistace (TMR) and tunneling anisotropic magnetoresistance (TAMR), respectively \cite{VonBergmann2012}. The blue circles correspond to lowest order magnetic 
Bragg peaks with primitive vectors ${\bf G}_1$ and ${\bf G}_2$, whereas the yellow circles are higher-order peaks located at $\pm{\bf G}_1\pm{\bf G}_2$ and $\pm {\bf G}_1 \mp {\bf G}_2$ around each atomic Bragg peak. The angle enclosed by the primitive vectors  differs from $90^\circ$, indicating that the magnetic periodicity is clearly distinct from a square lattice. There exists however a commensurability between the magnetic and the atomic periodicities given by the relation ${\bf g}_1 + {\bf g}_2 = 3 ({\bf G}_1 + {\bf G}_2)$.

The corresponding lattices in real space are shown in Fig.~\ref{fig1}a with ${\bf a}_i$ and ${\bf A}_i$ being primitive vectors of the atomic and magnetic lattice, respectively, with the standard relations ${\bf a}_i\cdot{\bf g}_j=2\pi\delta_{ij}$ and ${\bf A}_i\cdot{\bf G}_j=2\pi\delta_{ij}$ where $i = 1,2$. 
We find $|{\bf A}_1| = |{\bf A}_2| \approx 3.37 a \approx 9.15$ \AA\, and $\angle({\bf A}_1,{\bf A}_2) \approx 79.2^\circ$. 
The combinations ${\bf a}_\pm = {\bf a}_1 \pm {\bf a}_2$ are the primitive vectors 
of the equivalent centered rectangular lattice of the Fe monolayer. 
The vectors ${\bf a}_{+/-}$ define two important orthogonal directions that we denote in the following as {\it hard} and {\it soft}, respectively. The 3-fold commensurability is obtained along 
the hard direction ${\bf A}_1 + {\bf A}_2 = 3 {\bf a}_+$. Along the soft direction ${\bf a}_-$, the magnetic texture appears to be incommensurate with the atomic crystal. 
These findings are completely consistent with the previous report of Heinze {\it et al.} \cite{Heinze2011}. The orange lines in Fig.~\ref{fig1}a indicate the $3\times 5$ commensurate approximation of the magnetic unit cell that possesses a 5-fold commensurability along ${\bf a}_-$. Note that this commensurate unit cell is also distinct from a square, because its smaller vertex angle is approximately $87.8^\circ$ and thus smaller than $90^\circ$. 


\subsection{Theoretical prediction of a magnetic phason mode} 

\begin{figure}
\includegraphics[width=\columnwidth]{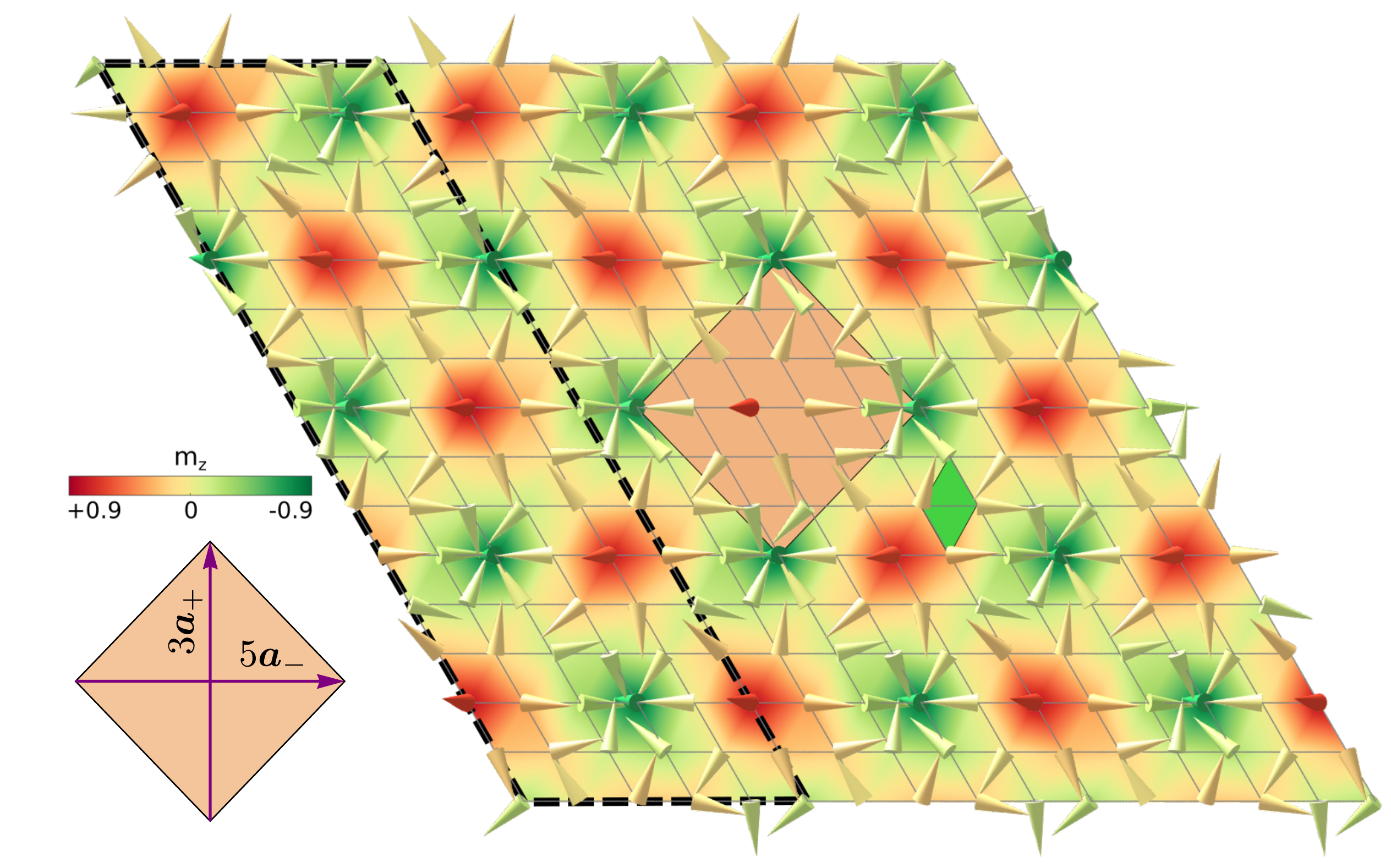} \hspace{1em}
\caption{Atomistic spin simulation of the $3\times 5$ commensurate magnetic texture. The green- and orange-shaded regions indicate the atomic unit cell and the commensurate approximation of the magnetic unit cell, cf.~Fig.~\ref{fig1}a. The region marked by a dashed line is the smallest common unit cell of the system. The supercell chosen for the simulation is three times larger and contains $15 \times 15$ magnetic atoms. The unit cell on the left-hand side indicates the ${\bf a}_\pm$ directions with the respective commensurabilities. 
}
\label{fig2}
\end{figure}

The uniaxial incommensurability identified above implies the existence of a gapless phason mode describing the sliding of the magnetic order along the soft ${\bf a}_- $ direction, which follows for infinitely large, clean systems from very general arguments 
\cite{ChaikinLubensky}. Nevertheless, in order to investigate theoretically the emergence of the phason mode, we performed atomistic spin simulations of the extended Heisenberg model put forward by Heinze {\it et al.} \cite{Heinze2011}, that reproduces important features of the observed magnetic texture in Fe/Ir(111). It consists of Hei\-sen\-berg, Dzyaloshinskii-Mo\-ri\-ya, biquadratic, and four-spin exchange interactions as well as uniaxial anisotropy, see {\bf Methods} for details. As an atomic simulation of an incommensurate magnetic texture is not feasible in a finite model, we limited ourselves to its $3\times 5$ commensurate approximation, see orange lines in Fig.~\ref{fig1}a. 
For that purpose, a supercell containing $15 \times 15$ magnetic moments positioned on the sites of the hexagonal atomic lattice was chosen; periodic boundary conditions were imposed that necessarily lead to the 
commensurate approximate texture. 
The resulting ground state is shown in Fig.~\ref{fig2} and reproduces the magnetic texture of Ref.~\cite{Heinze2011}.

\begin{figure*}
\includegraphics[width=\columnwidth]{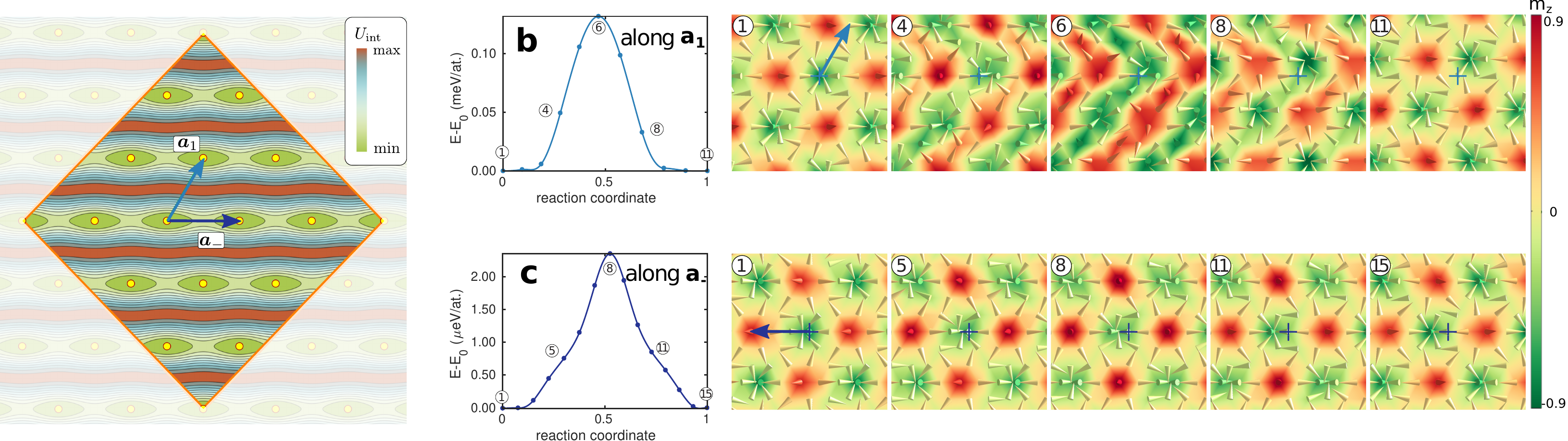}
\caption{{\bf a}. Illustrative sketch of the effective potential for the sliding phason mode for the $3 \times 5$ commensurate magnetic texture of Fig.~\ref{fig2}, see {\bf Methods} for details.
{\bf b} and {\bf c}. Geodesic nudged elastic band calculations for a shift of the magnetic texture along the ${\bf a}_1$ and ${\bf a}_-$ directions, respectively. The panels on the left- and right-hand side depict, respectively, the energy (blue dots and blue, interpolated curve) and magnetic configurations along the transition path. The cross markers on the left panels mark the initial position of an arbitrary spin initially pointing down.
The energy barriers for the two cases differ by two orders of magnitude. 
}
\label{fig3}
\end{figure*}

Generically, it is expected that the energetic cost for a shift of the magnetic texture increases with its commensurability, i.e., the potential barrier along the
hard ${\bf a}_+$ direction should be larger than along the soft ${\bf a}_-$ direction, see  Fig.~\ref{fig3}a. In order to confirm this expectation and to determine the effective potential for the phason mode, we performed geodesic nudged elastic band calculations, see {\bf Methods} for details. In particular, we considered a continuous shift of the texture along the two primitive vectors ${\bf a}_1$ and  ${\bf a}_-$ of the atomic lattice. We also checked that a shift along ${\bf a}_2 = {\bf a}_1 - {\bf a}_-$ yields the same results as the shift along ${\bf a}_1$. During these calculations, the minimum energy path between the initial and final states is determined. The evolution of the energy is shown in Fig.~\ref{fig3}b and c, together with examples of intermediate configurations of the magnetic texture. The explicit calculations show that the potential barrier for a shift along ${\bf a}_1$ is indeed two orders of magnitude larger than along the soft ${\bf a}_-$ direction. Moreover, the magnetic configuration is strongly deformed along the ${\bf a}_1$ path. This implies that a shift along the hard direction does not  involve only the phason excitation, but rather a superposition of additional modes, whose characteristics have been discussed in detail in Ref.~\cite{desplat2023eigenmodes}. In contrast, the intermediate configurations for the ${\bf a}_-$ path are basically undeformed during the shift and appear to be well captured by a single phason mode. 

The simulations confirm that the sliding of the $3\times 5$ commensurate magnetic texture along the soft ${\bf a}_-$ direction is well described by a phason that however possesses a small gap due to the five-fold commensurability. The magnetic texture realized in Fe/Ir(111) lacks this five-fold commensurability and is basically incommensurable along the soft direction, see Fig.~\ref{fig1}, so that in the experimental system the phason is expected to be gapless at least in the absence of disorder.

\subsection{Experimental detection of the phason in Fe/Ir(111) }

\begin{figure}[b]
\includegraphics[width=\columnwidth]{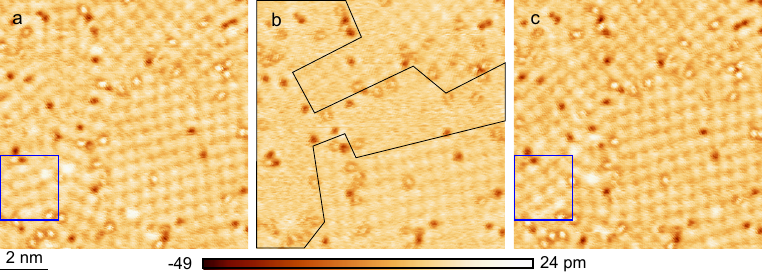}
\caption{{\bf a}. Topographic STM image of the magnetic texture observed at bias voltage $U$=10 mV and tunneling current $I$=10 nA. {\bf b}. The same area scanned at an elevated bias voltage of $U$=100 mV where the magnetic texture appears to be blurred within the region marked by black lines. {\bf c}. The subsequent image recorded at $U$=10 mV shows again a static texture where minor modifications with respect to panel {\bf a} can be spotted, e.g., in the area marked by the blue box. The legend indicates the colour code of the topographic height.}
\label{fig4}
\end{figure}

Fig.~\ref{fig4}a presents the magnetic texture on a larger scale recorded with STM topography using a non-magnetic tip at a bias voltage of $U$ = 10 mV. Here, the image contrast results only from the TAMR effect, which leads to a larger electronic density of states, i.e., higher $z$ position for in-plane than for out-of-plane magnetized Fe \cite{VonBergmann2012}. Several different rotational domains can be identified, as well as local defects that give
stronger contrasts. 

When measuring at a higher bias voltage of $U=$~100 mV, see Fig.~\ref{fig4}b, the spin texture becomes blurry or vanishes in some regions of the sample that are marked by black lines. This indicates that the magnetic texture in these regions is influenced by the action of the STM tip, i.e., it is either destroyed, melted or it is floating on the time scale of the STM experiment. Changing the voltage back to 10 mV subsequently reestablishes the static magnetic texture, see Fig.~\ref{fig4}c, albeit not in the exact same order as previously. Some rearrangements of the domains can be observed, e.g., by comparing the texture in the area marked by the blue boxes in panel a and c. The magnetic textures can thus be manipulated by the STM tip, as has been already demonstrated previously by von Bergmann {\it et al.} using short voltage pulses \cite{VonBergmann2014,VonBergmann2012}. We found however that most of the magnetic domains remained static in our sample even when scanned at elevated voltages. This suggests that the manipulation threshold depends on the local disorder configuration. 

\begin{figure}
\includegraphics[width=\columnwidth]{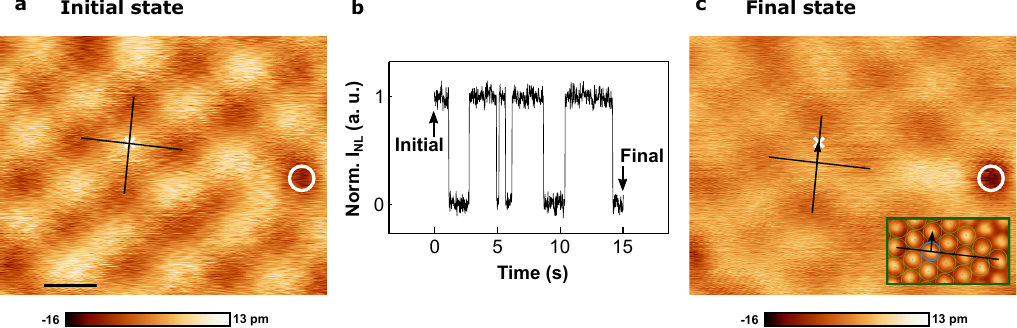}
\caption{{\bf a}. STM image of the initial state where the white circle marks a defect that serves as a reference for the scan area ($U$=10 mV, $I$=10 nA, scale bar: 0.5 nm).  The time trace of panel {\bf b} was recorded with the STM tip positioned at the location of the small white cross. The two black lines indicate the orientations ${\bf a}_{\pm}$ of the crystallographic unit cell, see Fig.~\ref{fig1}. {\bf b}. Time trace of the rectified current $I_{\rm NL}$ recorded at the position indicated by the white cross in {\bf a} and {\bf c} with $U$=80 mV, $I$=20 nA, and an additional ac voltage with amplitude $U_\mathrm{ac}$ = 27.2 mV at 2 GHz. {\bf c}. STM image of the final state ($U$=10 mV, $I$=10 nA). Compared to the initial state, the magnetic texture is laterally displaced along the soft crystallographic ${\bf a}_{-}$ direction. This is confirmed by the atomically resolved STM image of the crystal lattice shown in the inset.}
\label{fig5}
\end{figure}

%
%

A potential origin for the blurred images at larger bias voltage is the excitation of the phason mode of the magnetic texture induced by the STM tip. In order to test this hypothesis, we studied the dynamical behavior of the texture within this area in more detail. Fig.~\ref{fig5}a shows an STM scan of the investigated area where the white circle indicates a structural defect of the Fe film that is assumed to be immobile and serves as a benchmark. After placing the tip at the position of the white cross, the system was locally excited by increasing the bias voltage to $U=$ 80 mV and, in addition, applying a microwave ac voltage with amplitude $U_\mathrm{ac}$ = 27.2 mV at 2 GHz between the tip and the sample. For the calibration of the ac voltage and the rectification of the dc current $I_\mathrm{NL}$ see {\bf Methods}. Fig.~\ref{fig5}b shows the time trace of $I_\mathrm{NL}$ that exhibits a random telegraph noise (RTN) with a two-level switching characteristics. In the initial state, $I_\mathrm{NL}$ is high and switches several times from high to low and back within the 15 seconds of the measurement. At the end of the interval, $I_\mathrm{NL}$ is in the low-current state. Afterwards, an STM image of the final state was recorded and is presented in Fig.~\ref{fig5}c. Comparing the relative positions of TAMR contrast and the white cross, we observe that the magnetic texture between its initial and final state is shifted by approximately 0.25 nm, which is close to the atomic nearest-neighbour distance, $a$ = 2.715 \AA, of fcc Fe on Ir(111) \cite{Heinze2011}.
Moreover, inspection of the atomically resolved STM image in the inset of Fig.~\ref{fig5}c reveals that the lateral shift is along the incommensurate, soft ${\bf a}_-$ crystallographic direction, see Fig.~\ref{fig1}. 

The lateral shift is thus explained with the excitation of the sliding phason mode of the texture. It appears that this mode switches during the time trace of Fig.~\ref{fig5}b between two (meta-)stable states leading to a random telegraph noise. Note that the switching occurs on the time scale of seconds, and it is thus extremely slow compared to the applied GHz frequency. It is likely that the (meta-)stable states are generated by the presence of disorder, the various finite domains, as well as the concomitant domain walls present in the sample, which are clearly visible in Fig.~\ref{fig4}.

\subsection{Dependence of the switching rate}

\begin{figure}
\includegraphics[width=\columnwidth]{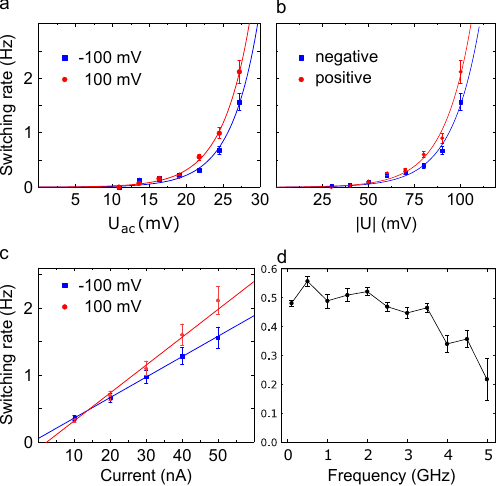}
\caption{Dependency of the switching rate extracted from the telegraph noise on the external parameters. (a) Dependence on the ac amplitude, blue (red) squares data represent taken at -100 mV (100 mV). Solid lines are guides to the eye ($U$=100 mV, $I$=50 nA, 2 GHz excitation). (b) Dependence of the dc voltage (symbols as in a, $I$=50 nA, $U_\mathrm{ac}$=27.2 mV, 2 GHz excitation). (c) Dependence on the current. Solid lines show the linear fit of the data ($U$=100 mV, $U_\mathrm{ac}$=27.2 mV, 2 GHz excitation). (d) Dependence on the micro-wave frequency ($U$=100 mV, $I$=15 nA, $U_\mathrm{ac}$=27.2 mV).}
\label{fig6}
\end{figure}

In order to address the mechanism of phason excitation, we measured the RTN varying the ac amplitude, dc voltage, current and micro-wave frequency (see Fig. \ref{fig6}). Note that prior to all frequency dependency measurement, the STM tip was positioned on the non-magnetic Ir(111) surface and the STM micro-wave transfer function was calibrated using the method described elsewhere \cite{Herve2015}. The micro-wave loss in the transmission line was then compensated at the output of the micro-wave generator such that the ac voltage at the tunneling junction is constant for every frequency. Clearly, the switching rate increases faster than linearly with both the ac amplitude and dc bias. 
Note that no ac voltage is necessary for switching, if the dc bias is high enough. The rate, however, scales linearly with tunneling current, indicating essentially a single-electron process. Interestingly, there is an asymmetry in the rates depending on the direction of the tunneling current. Always, a $\approx$30\% higher rate is observed for negative bias voltages, i.e. for tunneling of electrons from the tip into the sample.
A simple heating mechanism leading to activated transitions would show a super-linear behaviour with the electric power, i.e. it would steeply rise with both the sample bias and the tunneling current, which is not observed.
Instead, the data suggests that the maximal voltage, i.e. the dc and ac components combined, needs to reach a threshold for the transitions, speaking for an inelastic tunneling mechanism, which also agrees with the linear current relation.
As the tunneling current is spin-polarized due to the local magnetic order under the tip, there is a natural reason for an asymmetry of the rates, depending on the sign of the bias (and current) coming from excitations of magnetic structures via transfer of spin angular momentum between the tip and the sample. In the forward tunneling direction, minority electrons are required to inelastically excite the spin structure via magnon creation, while in the backward direction, majority electrons are needed such that the asymmetry reflects the spin-polarization of the tunneling current \cite{Balashov2008}. The observed asymmetry is in agreement with a switching mechanism via magnetic excitations with a minority spin polarization of Fe. The frequency dependence is, however, rather weak, with a tendency of reduced switching rates above a micro-wave frequency of 3 GHz. One may speculate that when the duration of the voltage maximum exceeding the inelastic energy threshold for switching becomes shorter than the switching time, switching becomes less probable. In this case, the data suggests switching times of the order of 0.3 ns.

\section{Discussion}

In this work, we presented evidence for a low-energy magnetic excitation, i.e., a sliding phason mode of the uniaxially incommensurate magnetic texture of Fe/Ir(111). The activation of this mode was interpreted to be at the origin of blurred STM images of the magnetic texture shown in Fig.~\ref{fig4}b. Further investigation on a specific magnetic domain revealed that the phason is pinned into two minima of an effective potential due to the presence of disorder in our sample. The STM tip was able to induce a switching between the two minima, leading to a random telegraph noise in the tunneling current with switching rates on the order of Hz. 

This evidence for a sliding phason mode raises interesting questions to be investigated in future research. In ideal samples with macroscopic domains without pinning by defects or boundaries, the phason could potentially lead to a floating phase where the magnetic texture is sliding along the incommensurate direction with no or a very 
low barrier, leading to liquid-like spatial correlations. Moreover, as a function of external parameters like temperature or magnetic field, it might be possible to induce a commensurate-incommensurate transition where either the magnetic texture becomes fully incommensurate or fully commensurate with the atomic crystal lattice. Finally, the presence of a sliding phason mode should give rise to anisotropic responses to an applied spin current that will strongly depend on its orientation with respect to the uniaxial incommensurability.

\section{Methods}
\label{sec:Methods}

\subsection{Scanning tunneling microscopy (STM)} 

Measurements were conducted with a home-built low-temperature (4.2 K) ultra-high vacuum ($p \approx 1 \times 10^{-10}$ mbar) STM paired with a Nanonis controller. The Ir(111) single-crystal was prepared by cycles of annealing ($\sim$1200 K) in an oxygen environment ($p \approx 1 \times 10^{-7}$ mbar) and high temperature flashing ($\sim$1700 K). The Fe film was deposited onto the clean Ir(111) surface using an electron-beam evaporator at slightly elevated temperature. STM tips were electrochemically etched from W wires and cleaned in-situ by flashing ($\sim$2500 K). Spin-polarized tips were prepared by depositing Fe on the W tips followed by soft annealing. The ac modulation was generated by a Rohde\,\&\,Schwarz SMB100A signal generator, mixed to the dc bias voltage by a Mini-circuits bias tee (ZX85-12G-S+) and sent to the STM tip. The ac voltage was modulated in amplitude (100\%) at a frequency below the cutoff frequency of the STM current amplifier (typically at 3 kHz). The supplementary current ($\Delta I$) induced by the ac modulation was detected with a lock-in amplifier by modulating the ac amplitude. 
It is the sum of 2 contributions: $\Delta I=I_{NL}+I_{h}$, where $I_{NL}$ corresponds to a rectification effect of the non-linear $I(U)$ characteristic when introducing micro-waves to the tunneling junction \cite{Herve2015,nunes1993,moult2011}. The second term, $I_{h}$, corresponds to the spin rectification effect \cite{Tulapurkar2005,Herve2019,Baumann2015}. 

\subsection{Atomistic spin simulations}

\paragraph{Heisenberg Hamiltonian}

We simulate a supercell comprised of $15 \times 15$ normalized magnetic moments $\mathbf{\hat{m}}_i$ on a hexagonal crystal lattice, with periodic boundary conditions. Simulations are carried out with Matjes \cite{matjes} and Spirit \cite{muller2019spirit} atomistic frameworks, and the results are cross-checked between the implementations.  We use the extended Heisenberg model Hamiltonian for Fe/Ir(111) from Ref. \onlinecite{Heinze2011} with parameters extracted from first-principles density functional theory (DFT) calculations,
 \begin{equation}\label{eq:hamil}
 \begin{split}
 \mathcal{H} & =   -  \sum_{ij} J_{ij} 
 \left(
 	 \mathbf{\hat{m}}_i  \cdot \mathbf{\hat{m}}_j
 \right)
  - \sum_{ij} \mathbf{D}_{ij} \cdot
 \left( 
 	\mathbf{\hat{m}}_i \times \mathbf{\hat{m}}_j
 \right) \\
 &  - K \sum_i m_{z,i}^2 
 + \mathcal{H}^{\mathrm{HOI}},\\
 \end{split}
\end{equation}
where $J_{ij}$ and  $D_{ij}$ are respectively the Heisenberg exchange and Dzyaloshinskii-Moryia interaction  coupling constants, and $K$ is the perpendicular magnetic anisotropy constant. $\mathcal{H}^{\mathrm{HOI}}$ contains higher-order interactions (HOI) of the form,
 \begin{equation}\label{eq:hamil_HOI}
  \begin{split}
 \mathcal{H}^{\mathrm{HOI}}  = & - \sum_{ij} B_{ij}
 \left(
 	\mathbf{\hat{m}}_i \cdot \mathbf{\hat{m}}_j
 \right)^2 \\
 & - \sum_{ijkl} Q_{ijkl} 
 [
 	\left(
 		\mathbf{\hat{m}}_i \cdot \mathbf{\hat{m}}_j
 	\right)
 	\left(
 		\mathbf{\hat{m}}_k \cdot \mathbf{\hat{m}}_l
 	\right)	\\
 	& +
 	\left(
 		\mathbf{\hat{m}}_i \cdot \mathbf{\hat{m}}_l
 	\right)
 	\left(
 		\mathbf{\hat{m}}_j \cdot \mathbf{\hat{m}}_k
 	\right)	\\
 	& -
 	\left(
 		\mathbf{\hat{m}}_i \cdot \mathbf{\hat{m}}_k
 	\right)
 	\left(
 		\mathbf{\hat{m}}_j \cdot \mathbf{\hat{m}}_l
 	\right)	
  ],
\end{split}
\end{equation}
in which  $B_{ij}$, and $Q_{ijkl}$  denote the biquadratic, and four-spin interaction coupling constants. 

The relaxed ground state shown in Fig. \ref{fig2} has an energy of -9.97 meV/at. with respect to the ferromagnetic state~\cite{desplat2023eigenmodes}, which is comparable to the value of -7 meV/at. given in Ref. ~\onlinecite{Heinze2011} for the DFT unit cell.

\paragraph{GNEB simulations}
Minimum energy paths for the translation of the  magnetic texture along the different axes discussed in the main text are computed via the geodesic nudged elastic band  (GNEB) scheme~\cite{bessarab2015method}. 
The method minimizes the energy of magnetic configurations belonging to an effective elastic band within the multidimensional energy landscape by taking into account the curvature of configuration space. The energy minimization is carried out following negative gradient lines. The saddle point configuration at the top of the barrier is precisely relaxed with a climbing image scheme.


We find energy barriers $\Delta E_{\mathbf{a}_{1,2}} =   0.13 $~meV/at. for translation along $\mathbf{a}_{1,2}$, and $\Delta  E_{\mathbf{a}_-} =  2.7$~$\mu$eV/at. for translation along $\mathbf{a}_-$, i.e., the soft direction.


\subsection{Effective phason potential for a $3\times 5$ commensurate magnetic texture}

We elaborate on the construction of the effective phason potential expected for a $3\times 5$ commensurate magnetic texture that is sketched in Fig.~\ref{fig3}a. As discussed in the context of Fig.~\ref{fig1},  the reciprocal lattice vectors of the magnetic lattice $\bm{G}_i$ and the crystal lattice $\bm{g}_i$ satisfy the relation $\bm{g}_1 + \bm{g}_2 = 3(\bm{G}_1 + \bm{G}_2)$. Only for the $3\times 5$ commensurate approximation we have the additional relation $\bm{g}_1 - \bm{g}_2 = 5(\bm{G}_1 - \bm{G}_2)$, and thus solving for $\bm{g}_i$
\begin{align}
\bm{g}_1 = 4 \bm{G}_1 - \bm{G}_2, \quad
\bm{g}_2 = - \bm{G}_1 + 4\bm{G}_2.
\end{align}
In general, it is expected that the size of the Fourier components of the phason potential decreases with decreasing commensurability.
Restricting ourselves to the Fourier components with the most commensurate wavevectors, we approximated
\begin{align}
U_{\rm int}(\bm{r}) \approx - U_3 \cos((\bm{g}_1 + \bm{g}_2)\bm{r}) - U_4 ( \cos(\bm{g}_1 \bm{r}) + \cos(\bm{g}_2 \bm{r}))
\end{align}
where $U_{3/4} > 0$ and $U_4/U_3 \ll 1$ in the sketch of Fig.~\ref{fig3}a. The phason potential obtained in the simulations, see in particular Fig.~\ref{fig3}c, deviates from a single cosine which, in principle, can be accounted for by including higher-order Fourier components.



\section{Acknowledgements}
B.D. and L.D. thank Sebastian Meyer, Joo-Von Kim for useful discussions, and Markus Hoffmann and Gideon Müller for their help with Spirit. H.H. acknowledge funding by the Alexander-von-Humboldt Foundation. W.W. acknowledges funding by Deutsche Forschungsgemeinschaft (DFG) with grant Wu 394/15-1 and through CRC TRR 288 - 422213477 “ElastoQMat,” project B06. L.D. acknowledges funding by the University of Li{\`e}ge under Special Funds for Research, IPD-STEMA Programme. M.G. acknlowedges funding by the DFG through CRC TRR 288 - 422213477 (project A11) and through SPP 2137 (project 403030645).

%

\end{document}